\documentclass[reprint,twocolumn,showpacs,amsmath,amssymb,superscriptaddress,aps,prl]{revtex4-1}

\usepackage{graphicx}
\usepackage[english]{babel}

\usepackage{color}
\newcommand{\red}[1]{{#1}}

\begin{document}

\title{Low-noise amplification and frequency conversion with a multiport microwave optomechanical device}
\author{C. F. Ockeloen-Korppi}
\author{E. Damsk\"agg}
\author{J.-M. Pirkkalainen}
\affiliation{Department of Applied Physics, Aalto University, 
PO Box 11100, FI-00076 Aalto, Finland}
\author{T. T. Heikkil\"{a}}
\author{F. Massel}
\affiliation{Department of Physics, Nanoscience Center, University of Jyv\"askyl\"a, PO Box 35 (YFL), FI-40014 University of Jyv\"askyl\"a, Finland}
\author{M. A. Sillanp\"{a}\"{a}}
\affiliation{Department of Applied Physics, Aalto University, 
PO Box 11100, FI-00076 Aalto, Finland}

\date{\today}

\begin{abstract}
\red{High-gain amplifiers of electromagnetic signals operating near the quantum limit are crucial for quantum information systems and ultrasensitive quantum measurements. However, the existing techniques have a limited gain-bandwidth product and only operate with weak input signals. Here we demonstrate a two-port optomechanical scheme for amplification and routing of microwave signals, a system that simultaneously performs high-gain amplification and frequency conversion in the quantum regime. Our amplifier, implemented in a two-cavity microwave optomechanical device, shows 41 dB of gain and has a high dynamic range, handling input signals up to $10^{13}$ photons per second, three orders of magnitude more than corresponding Josephson parametric amplifiers. We show that although the active medium, the mechanical resonator, is at a high temperature far from the quantum limit, only 4.6 quanta of noise is added to the input signal. Our method can be readily applied to a wide variety of optomechanical systems, including hybrid optical-microwave systems, creating a universal hub for signals at the quantum level.}
\end{abstract}

\maketitle

Recent advances in near-quantum limited amplifiers \red{in the microwave-frequency regime} have lead to breakthroughs in the understanding of quantum measurement processes \cite{Clerk:2010dh,VijaySiddiqi2011,Teufel2011b,Wallraff2013Teleport}, and are necessary in quantum error correction and feedback \cite{Siddiqi2012,Martinis2015Error,Kippenberg2015FB}. Of particular interest for most applications are phase-insensitive linear amplifiers, which provide a faithful reconstruction of both quadratures of the input. Such amplifiers are bound by the standard quantum limit (SQL), which states that at high gain at least half an energy quantum of noise is added to the input signal \cite{Caves1982}. This limit has been approached with Josephson parametric amplifiers~\cite{Yurke1989,Castellanos-BeltranLehnert2008,BergealDevoret2010,Lahteenmaki2012,Wallraff2modent,MacklinSiddiqi2015}, however, such amplifiers are limited to relatively \red{weak input signals. Optomechanical amplifiers or detectors}, utilizing the interaction between electromagnetic waves and a mechanical resonator \red{inside a cavity}~\cite{AspelmeyerMarquardt2014}, have been demonstrated in the microwave and optical regime~\cite{MasselSillanpaa2011,McRaeBowen2012,Minnesota2012OptoAmp,Polzik2014Amp}, but existing techniques suffer from a limited gain and bandwidth as well as noise levels well above the SQL.

In this paper, we demonstrate a two-port optomechanical device, motivated by the proposal in Ref.~\cite{MetelmannClerk2014}. It consists of two electromagnetic cavities with different resonant frequencies, simultaneously coupled \cite{Painter2010Laser,DongWang2012,HillPainter2012} to a single mechanical resonator. 
In the presence of appropriately chosen external pump tones, the mechanical resonator mediates interaction between the cavities, enabling strong amplification of a signal reflected from one of the cavities. Moreover, the scheme supports frequency-converting amplification, where a signal incident in one cavity can irradiate out from the other cavity at a completely different frequency, while being amplified at the same time. As we show theoretically, our scheme can reach the SQL for both of these processes.
Unlike existing optomechanical amplifiers~\cite{MasselSillanpaa2011,McRaeBowen2012}, the bandwidth of amplification in our scheme can be increased up to the cavity linewidth, and the product of gain and bandwidth has no fundamental limit. \red{Remarkably, we show in experiment that the quantum limit can be closely approached even at a high temperature where the mode occupation numbers $\gg 1$, which allows for an interpretion in terms of reservoir engineering \cite{Zoller1996BathEng,Siddiqi2012BathEng,Devoret2013ResEng,WangClerk2013,MetelmannClerk2014}.}

In a microwave-frequency optomechanical \cite{Lehnert2008Nph} experiment, we obtain a gain of 41~dB with a gain-bandwidth product of 137~kHz, while adding only 4 quanta of noise above the SQL. Additionally, we show how the multimode system can act as a spectrally pure microwave source in the regime of self-oscillations. Finally, using an alternate pump scheme, our system also supports frequency conversion without amplification, similar to previous experiments in a range of systems~\cite{DongWang2012,HillPainter2012,AndrewsLehnert2014,LecocqTeufel2016}. With that method, we observe near-unity conversion efficiency, wide bandwidth and added noise on the single photon level. 

\begin{figure*}
\includegraphics[width= 0.95 \textwidth]{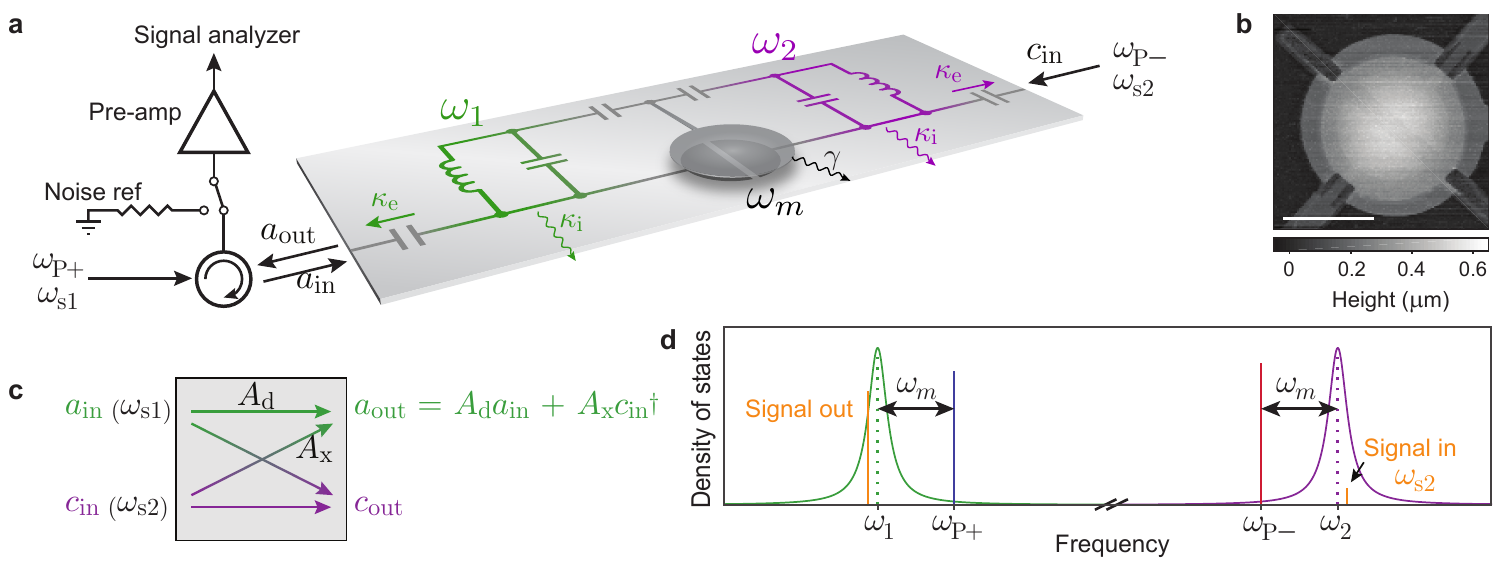}
\caption{Experimental setup. 
a) Schematic of our device, showing two LC microwave cavities ($\omega_1$, $\omega_2$) both coupled to a central mechanical drum resonator ($\omega_m$) as well as an individual feedline. Signals ($\omega_\mathrm{s1},\omega_\mathrm{s2}$) and pumps ($\omega_\mathrm{P+},\omega_\mathrm{P-}$) are fed to the cavities as shown. The output of cavity 1 is pre-amplified and measured with a signal analyzer.
b) Atomic force micrograph of the drum resonator. Scale bar is $10~\mu$m long.
c) Conceptual two-port amplifier, exhibiting both direct ($A_\text{d}$) and cross ($A_\text{x}$) gain. 
d) Schematic representation of the cavity modes and pump frequencies used to realize a two-port amplifier. As an example, an input signal with frequency $\omega_\mathrm{s2}$ is injected to cavity 2, with the frequency-converted and amplified output emerging from cavity 1.
}
\label{fig:setup}
\end{figure*}

Our setup is shown in figure~\ref{fig:setup}, and discussed in detail in the supplementary information~\cite{SI}. At the heart of our device is a suspended aluminium drum resonator \cite{Teufel2011b}, with a resonant frequency of $\omega_m = 2\pi\times 8.3$~MHz and a line width of $\gamma = 2\pi\times100$~Hz, fabricated on a quartz substrate. The drum is surrounded by two inductor-capacitor (LC) cavities, with resonant frequencies $\omega_1 = 2\pi\times7.0$~GHz and $\omega_2 = 2\pi\times8.4$~GHz, respectively. The mechanical resonator is suspended over two electrodes to form a variable capacitance in each cavity, simultaneously coupling the mechanical motion to both cavities. Each cavity is capacitively coupled to an individual transmission line with a strong coupling rate $\kappa_\text{e} = 2\pi\times4.8$~MHz compared to the internal loss rate $\kappa_\text{i} = 2\pi\times0.50$~MHz. The total cavity line width is $\kappa = \kappa_\text{i} + \kappa_\text{e}$.

The two-mode amplifier is created by injecting two pump tones (figure~\ref{fig:setup}d), one at the blue mechanical sideband of cavity 1 ($\omega_{P+} = \omega_1 + \omega_m$), and the other at the red sideband of cavity 2 ($\omega_{P-} = \omega_2 - \omega_m$). \red{Quantum-limited amplification} in related multi-cavity configurations have been theoretically proposed~\cite{MetelmannClerk2014,NunnenkampKippenberg2014,MetelmannClerk2015}, involving artificially increased mechanical damping. However, we show that the performance of the amplifier under proper pumping conditions in fact improves with reducing $\gamma$. Moreover, we introduce the possibility for frequency-converting amplification, \red{in addition to unity-gain frequency conversion \cite{DongWang2012,HillPainter2012,AndrewsLehnert2014,LecocqTeufel2016}.}

In the presence of strong pump tones, the optomechanical interaction is effectively linear and enhanced by the cavity, with coupling strengths $G_+ = \sqrt{n_1}g_1$ and $G_- = \sqrt{n_2} g_2$, where $g_i$ is the single-photon coupling strength and $n_i$ the photon occupation of cavity $i$. The effective Hamiltonian describing the coupling between the subsystems is
\begin{equation}
H_I = (G_- c^\dagger + G_+ a)b + \text{h.c.},
\end{equation}
($\hbar=1$ hereafter) where the operators $a,a^\dagger$ and $c,c^\dagger$ represent cavity 1 and 2, respectively, and $b,b^\dagger$ represent the mechanical mode. 
$H_I$ can be interpreted as a two-step process composed of nondegenerate parametric amplification \red{\cite{Louisell1961,Clerk:2010dh} between cavity 1 and the mechanics (term $G_+a b + \text{h.c.}$), followed by a beam splitter between the mechanics and cavity 2 (term $G_- c^\dagger b + \text{h.c.}$) which acts as a means to transfer the amplification to cavity 2.}

\begin{figure}
\includegraphics[width=0.99 \columnwidth]{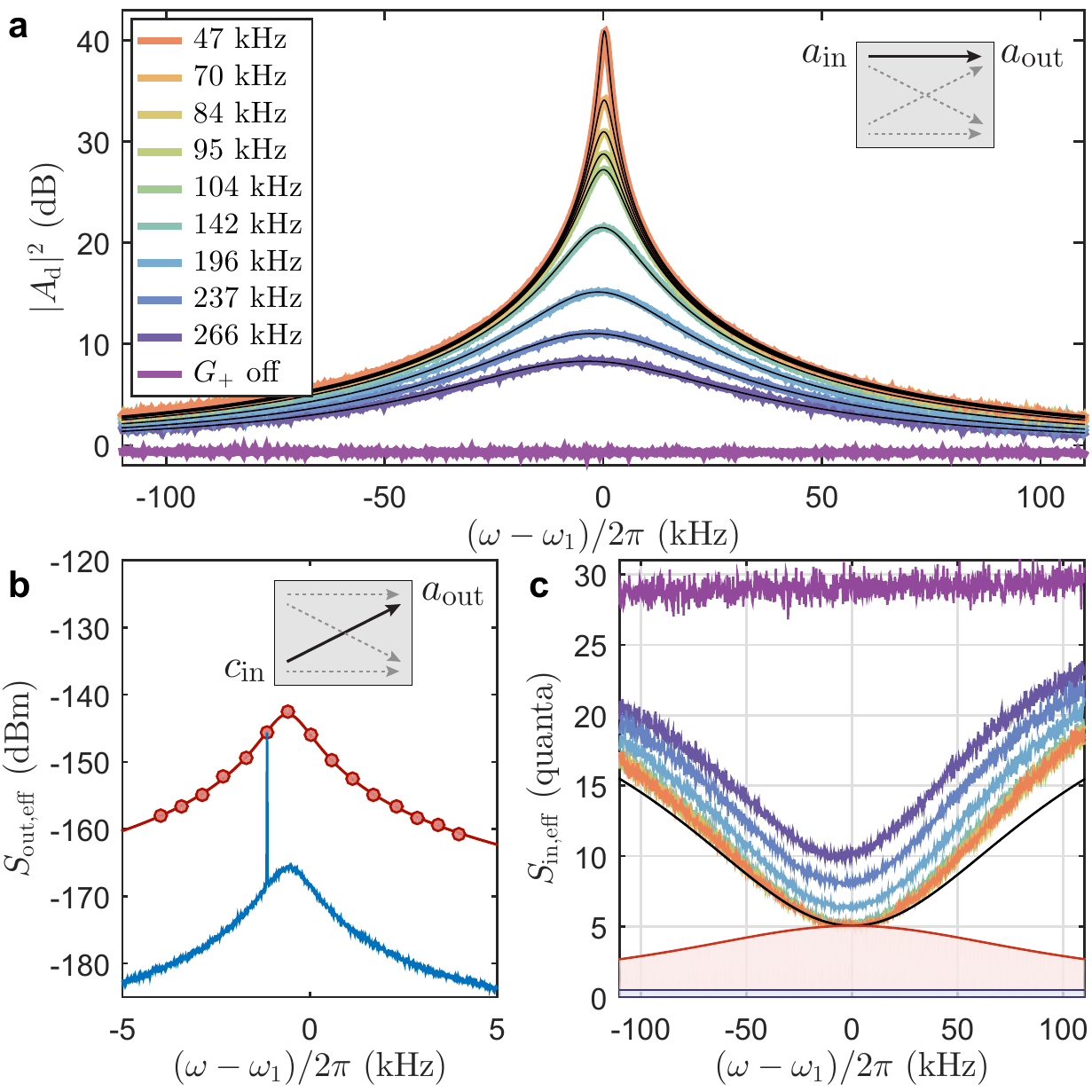}
\caption{Two-mode amplifier performance. 
a) Direct gain $|A_\text{d}|^2$ versus signal frequency for fixed $G_-$ and various values of $G_+$ (colored lines, legend shows $\mathcal{G}/2\pi$) together with theory fits (black lines). Inset: amplifier configuration, the signal is input to cavity 1.
b) Example output spectrum of cavity 1 (blue line), showing high-gain frequency conversion of a weak signal injected in cavity 2 (narrow peak). The peak height at different frequencies (circles) agrees with the fitted model (red line). Inset: amplifier configuration.
c) Effective input-referred noise for the same $G_-$,$G_+$ as in panel a (colored lines). The theory model (black line) is plotted only for the highest gain. The blue shaded area shows the input noise of one half quanta at each input, and the red shaded area shows the modelled added noise $S_\text{add}$.}
\label{fig:gainandnoise}
\end{figure}

Using input-output theory, and assuming $G_- \gtrsim G_+$, we find the system behaves as a two-port phase-insensitive linear amplifier, as depicted schematically in figure~\ref{fig:setup}c. The output field $a_\text{out}$ of cavity 1 has the from
\begin{equation}\label{eq:inout}
a_\text{out} = A_\text{d} a_\text{in} + A_\text{x} c_\text{in}^\dagger + F,
\end{equation}
and similar for $c_\text{out}$ of cavity 2. Here $A_\text{d}$ is the direct gain of signals $a_\text{in}$ incident on cavity 1, and $A_\text{x}$ is the cross (frequency-converting) gain of signals incident on cavity 2. Operator $F$ 
describes the added noise due to the internal modes of the device.
High-gain amplification is achieved with nearly equal coupling strengths, which we parametrize by $\mathcal{G}^2 = (G_-^2 - G_+^2)$. For $\mathcal{G}^2 \ll G_-^2$, the direct and cross gains are approximately equal, with peak values on resonance
\begin{equation}\label{eq:peakgain}
|A_\text{d}|^2 \approx |A_\text{x}|^2 \approx \left|2\frac{\kappa_\text{e}}{\kappa} \frac{4 G_-^2/\kappa}{\gamma_\text{eff}} \right|^2,
\end{equation}
where $\gamma_\text{eff} = \gamma + 4\mathcal{G}^2/\kappa$ is the effective damping of the mechanical oscillator. \red{Similar to optomechanical amplifiers powered by a single blue-detuned pump \cite{MasselSillanpaa2011}, the amplification bandwidth is associated with the effective mechanical linewidth $\gamma_\text{eff}$. In the present case, however, the parametric instability threshold is never reached} since $\gamma_\text{eff}>0$ (see~\cite{WangClerk2013}). Moreover, the gain-bandwidth product $\text{GBW} = |A_\text{d}|\gamma_\text{eff}$ is determined by $G_-$, and is not fundamentally limited. 

To characterize the noise performance, we calculate the expected output noise power-spectral density (PSD) $S_\text{out} = \frac{1}{2}\langle a_\text{out}^\dagger a_\text{out} + a_\text{out}a_\text{out}^\dagger \rangle $ from equation~\ref{eq:inout} (and similar for $c_\text{out}$). The input-referred added noise is then calculated as $S_\text{add} = (S_\text{out} - S_\text{in})/|A|^2$, where $A$ is $A_\text{d}$ or $A_\text{x}$ for direct and frequency-converting amplification, respectively, and $S_\text{in}$ is the noise of the input state. We find that for large $G_-$ and $G_+$, strong external coupling $\kappa_\text{e} \gg \kappa_\text{i},\gamma$, the added noise approaches the quantum limit of one half quantum for both direct and frequency-converting amplification \cite{SI}. These conditions are available in realizations using either microwave or optical cavities. \red{In a stark contrast to the regular nondegenerate parametric amplifier \cite{Louisell1961}, the quantum limit can practically be reached although the environment is at a very high temperature.}


In experiment, we measure the performance of our amplifier in a cryogenic environment at a base temperature of 7~mK. We inject pumps and signals in both cavities, while measuring the output of cavity 1. Figure~\ref{fig:gainandnoise}a shows the direct gain $|A_\text{d}|^2$, where the signal was injected into cavity 1. Data is shown as a function of signal frequency for $G_- = 2\pi\times 355$~kHz and several values of $G_+$ up to $G_+ = 0.99 \, G_-$, corresponding to the highest gain. We achieve a maximum gain of $|A_\text{d}|^2 = 41$~dB with a 3~dB-bandwidth of $\gamma_\text{eff}=1.2$~kHz, resulting in GBW = 137~kHz. The data is in excellent agreement with fits to our model (see supplementary information). Figure~\ref{fig:gainandnoise}b shows an example of frequency-converting amplification. Shown is the output spectrum of cavity 1, while a weak sinusoidal signal was injected into cavity 2 close to resonance, for similar pump strengths as in panel a. The converted output signal is visible as a narrow peak. The peak output was measured for several frequencies (circles), and the peak gain was $|A_\text{x}|^2 \approx 26$~dB. 

To accurately quantify the noise performance we calibrate the total system gain with two independent methods (see also supplementary information). First, we compare to a resistor with known 2.9~K thermal noise at the output side of the sample (figure~\ref{fig:setup}a). Second, in a subsequent cooldown, we verified the calibration with a tunable noise source at the input side of our sample. Using the latter method, we do not need to know the cable attenuation between the sample and the pre-amplifier. 
Figure~\ref{fig:gainandnoise}c shows the effective input-referred noise spectrum $S_\text{in,eff} = S_\text{out,eff}/|A_\text{d}|^2$ with no signal input, expressed as number of quanta per unit bandwidth, regarding cavity 1 as the input port. $S_\text{in,eff}$ is the total system noise: it includes the input vacuum noise ($1/2$ quantum, blue shaded area), the added noise $S_\text{add}$ of the mechanical amplifier (red shaded area), as well as the output technical noise $S_\text{H,eff}$ added by all further amplification stages. The latter dominates off-resonance, but is negligible at high gain. The measured effective noise therefore reduces at increasing gain, saturating at $S_\text{in,eff} \simeq 5$~quanta for $|A_\text{d}|^2 \gtrsim 20$~dB. This corresponds to an added noise of $S_\text{add} = 4.6 \pm 1.0$ quanta at the highest gain measurement. The same result applies to frequency-converting amplification (with cavity 2 as the input port), since at high gain $|A_\text{x}| \approx |A_\text{d}|$. The uncertainty of $S_\text{add}$ is dominated by the residual power calibration uncertainty. 

\begin{figure}
\includegraphics[width= 0.99 \columnwidth]{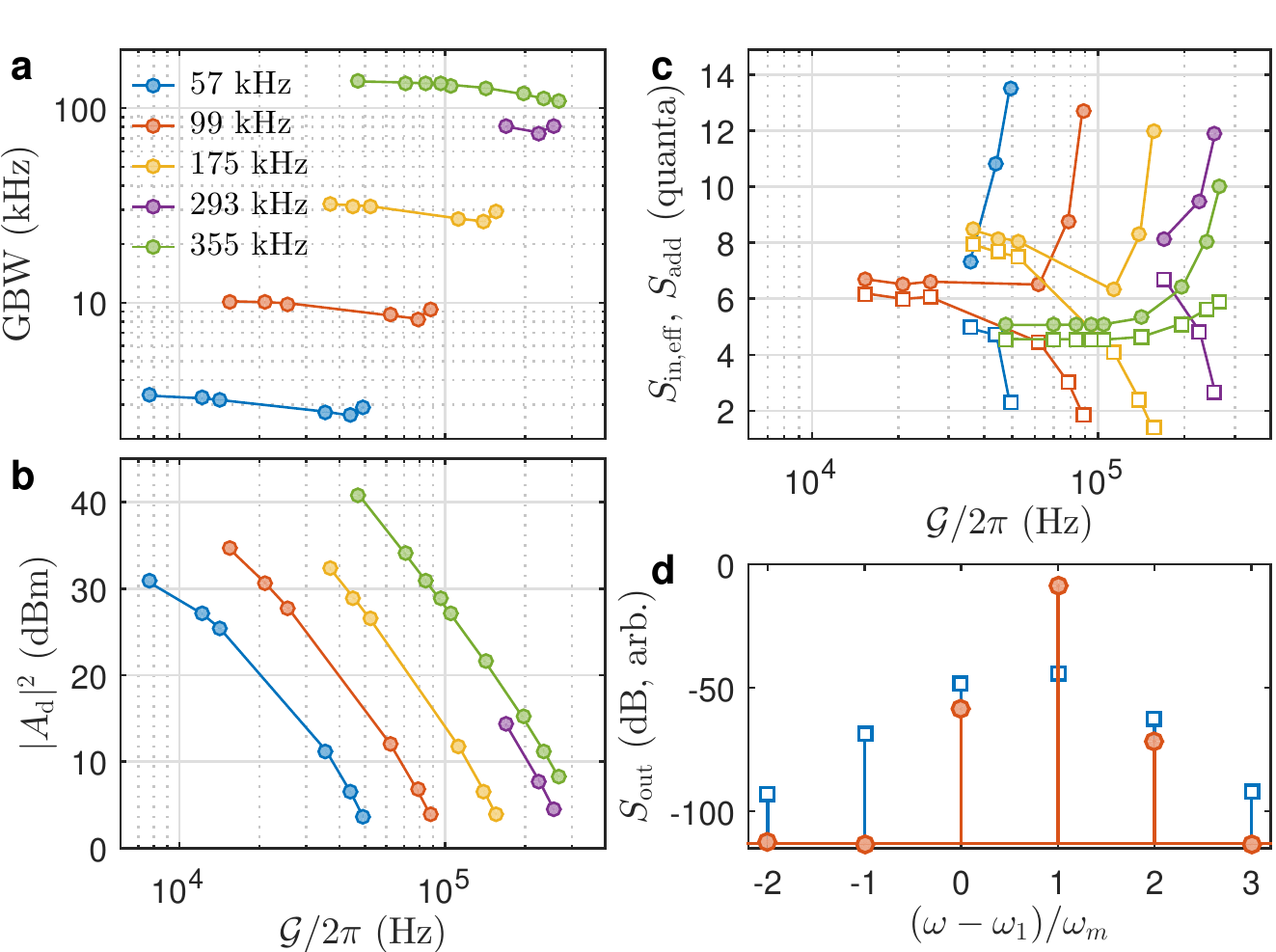}
\caption{Amplifier performance and oscillations. a) Gain-bandwidth product for several values of $G_-$ (legends show $G_-/2\pi$) as a function of $\mathcal{G}$. b) Peak direct gain and c) effective input-referred noise (closed circles) and added amplifier noise $S_\text{add}$ (open squares) on resonance, for the same parameters as panel a. d) Harmonic spectra of oscillations in the instability regime without (open squares) and with (closed circles) the red-detuned pump at $\omega_\text{P-}$ enabled. For the latter case, each harmonic component was measured separately and the horizontal line shows the measurement noise floor.}
\label{fig:fitresults}
\end{figure}

In figure~\ref{fig:fitresults}a-c we summarize the performance of our amplifier for a wide range of pump powers. For fixed $G_-$, the gain-bandwidth product is approximately independent of $\mathcal{G}$, as expected from our model. The highest GBW = 137~kHz as well as the highest absolute gain are achieved at higest $G_-$, which was in turn limited only by the experimentally available pump power. Figure~\ref{fig:fitresults}c shows the total input-referred noise $S_\text{in,eff}$ on resonance, as well as the added noise $S_\text{add}$ of our amplifier.  At low pump powers, and thus low gain, we observe added noise below 2 quanta, but for high pump powers $S_\text{add}$ increases. Comparing to our theory model, we find that the noise performance is well described by heating of the mechanical environment at high pump powers, up to an effective temperature $T_\text{env}$ corresponding to $n^T_m = k_B T_\text{env}/\hbar\omega_m \approx 5\times10^3$ quanta. These results are consistent with the heating we observe with standard optomechanical cooling measurements (see SI~\cite{SI}). While heating processes limit the noise performance of the current experiment, they do not pose a fundamental limit on our scheme. The heating could be reduced by increased optomechanical couplings $g_1$ and $g_2$, improved mechanical and cavity quality factors, and further depends on the details of device fabrication. 

In addition to operating the pump frequencies at sideband co-resonance, our amplifier can be tuned over the cavity linewidth $\kappa$ by shifting the pump frequencies. We have measured high-gain ($|A_\text{d}|^2 > 26~$dB) and low-noise ($S_\text{in,eff} \approx 5~$quanta) amplification over a tuning range of 2~MHz. Further tuning could be achieved with tunable microwave cavities \cite{Lehnert2015}. Free-space \red{optical cavities}, which have been used in optomechanical systems, directly provide tuning over a wide range.

In contrast to existing amplifiers operating near the quantum limit, our amplifier can handle large signal levels. With a gain of 23~dB our and similar pump powers as in figure~\ref{fig:gainandnoise}, our amplifier remains stable up to an input power of $-69~$dBm, or $3\times10^{13}$ photons/s. This is $30~$dB higher than reported in Josephson parametric amplifiers~\cite{MacklinSiddiqi2015}, and corresponds to a \red{very large} dynamic range of $127~$dB in a 1~Hz measurement bandwidth. In these large-signal measurements, the signal-to-noise ratio (SNR) was $>100~$dB in a 1~Hz measurement bandwidth, limited by the phase noise of our signal generator. Since we see no evidence of increased noise towards higher input powers, we believe the SNR is equal to the dynamic range, \red{which is five orders of magnitude higher than in typical Josephson parametric amplifiers \cite{Castellanos-BeltranLehnert2008}, and comparable to the highest values obtained with nearly quantum-limited microwave measurement systems \cite{Day2012KinIndAmp}}.

When increasing $G_+$ beyond the stability requirement $\gamma_\text{eff} > 0$, the system undergoes a lasing transition to self-sustained oscillations, which can be a source of spectrally pure electromagnetic radiation~\cite{NunnenkampKippenberg2014}. Figure~\ref{fig:fitresults}d shows the measured harmonic spectrum of such oscillations for the two-cavity case (both pumps $\omega_\text{P+}$ and $\omega_\text{P-}$ on) and for the case of a single pump $\omega_\text{P+}$. Whereas the single pump case shows many harmonic components, in the two-cavity case we observe only a single sideband at $\pm \omega_m$ around the pump frequency $\omega_\text{P+}$. The absence of higher sidebands demonstrates that the mechanical resonator has pure sinusoidal oscillations, which can allow for applications as a source of a clean clock signal.

In a third experiment, we demonstrate coherent frequency conversion of microwave signals without amplification, a frequency-converting analog to an optical beam splitter. This method has been previously demonstrated with optical frequencies~\cite{DongWang2012,HillPainter2012}, hybrid microwave-optical systems~\cite{AndrewsLehnert2014}, and very recently in a system similar to ours~\cite{LecocqTeufel2016}. Both cavities are pumped at the red mechanical side-band, with pump frequencies $\omega_{\text{P}i} = \omega_i - \omega_m$ for cavity $i=1,2$, respectively. A weak input signal with frequency $\omega_s$ is injected into cavity 2. The converted signal appears at the output of cavity 1, with a frequency $\omega_s' = \omega_s - \omega_{\text{P}2} + \omega_{\text{P}1}$. The internal conversion efficiency between the two cavities is~\cite{DongWang2012}
\begin{equation}\label{eq:freqconv_etaint}
\eta_\text{int} = \frac{4 G_1^2 G_2^2}{(G_1^2 + G_2^2 + \frac{\gamma\kappa}{4})^2},
\end{equation}
which approaches unity for $G_1^2 = G_2^2 \gg \frac{\gamma\kappa}{4}$. Here, $G_i = \sqrt{n_i}g_i$. The total conversion efficiency from input to output is $\eta = \eta_\text{int} \kappa_\text{e}^2 / \kappa^2$.

\begin{figure}
\includegraphics[width=0.99 \columnwidth]{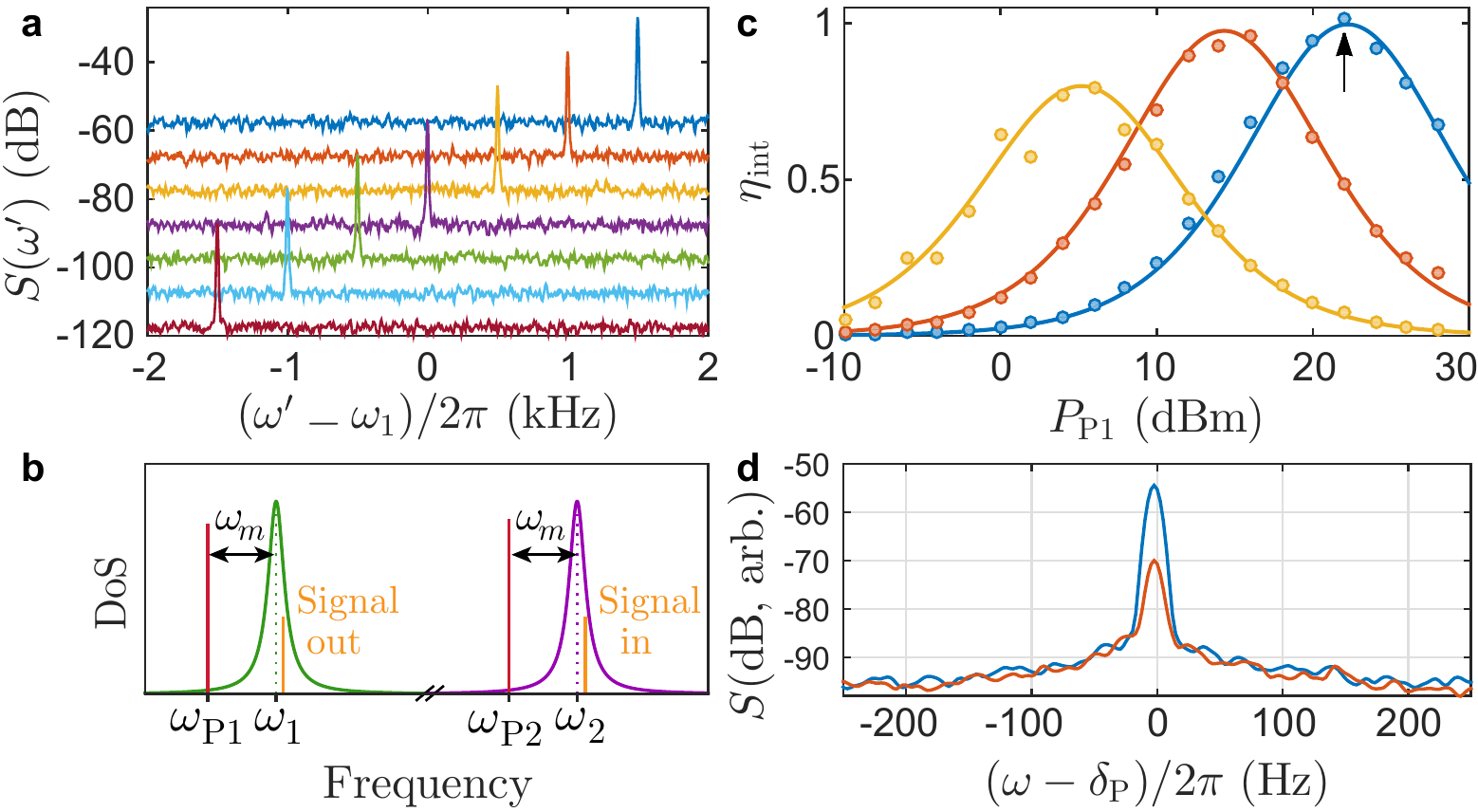}
\caption{Frequency conversion without amplification. a) Output spectrum $S(\omega')$ (arbitrary units, each curve offset by 10~dB) showing the frequency-converted output signal $\omega_s' \approx \omega_1$ for several input frequencies $\omega_s \approx \omega_2$. b) Pump and signal configuration. c) Conversion efficiency on resonance ($\omega_s = \omega_2$) along with fitted theory (see text). Arrow indicates data shown in panel a. d) Coherence of conversion. Shown are spectra of the mixed-down pump tone $\delta_\text{P}$ (blue line) and the destructive interference of $\delta_\text{P}$ and $\delta_s$ (red line).}
\label{fig:freqconv}
\end{figure}

The experimental results are shown in figure~\ref{fig:freqconv}. Panel a shows the output power spectral density $S_\text{out,eff}(\omega')$ for several input frequencies close to $\omega_2$. The output signal is visible as a narrow peak.  The conversion has a bandwidth of 50~kHz, corresponding to the mechanical line-width broadened by optomechanical cooling by both pumps. Panel c shows the maximum internal conversion efficiency at $\omega_s = \omega_2$ as function of output-cavity pump power $P_{\text{P}1}$, for various input-cavity pump powers $P_{\text{P}2}$, together with a global fit of all the data to equation~\ref{eq:freqconv_etaint}. We obtain a peak internal efficiency of $\eta_\text{int} > 0.99$, and the total efficiency is only limited by the cavity coupling $\eta \approx \kappa_\text{e}^2/\kappa^2 \gtrsim 0.82$. Similar to the two-port amplifier, we observe an increase in noise due to heating of the mechanics at the highest pump powers, resulting in an added noise of $3.9$ quanta at near-unity conversion efficiency. Using lower pump powers, we obtain an added noise of $1.4$ quanta, while retaining a conversion efficiency of $\eta_\text{int} = 0.95$. In optical and hybrid systems much lower efficiencies have been reached, mostly limited by $\kappa_\text{e}$~\cite{DongWang2012,HillPainter2012,AndrewsLehnert2014}. To verify coherence of the conversion process, we externally mix the input and output signals, generating their difference frequency $\delta_s = \omega_s - \omega_s'$. Similarly, we mix together the two probe tones to generate $\delta_\text{P} = \omega_{\text{P}2} -\omega_{\text{P}1}$. We then combine $\delta_s$ and $\delta_\text{P}$ on a resistive adder, and with appropriately adjusted relative phase and amplitude they destructively interfere by 15~dB (figure~\ref{fig:freqconv}d). We thus demonstrate coherent frequency conversion near the quantum limit.

Our concept of a two-port optomechanical phase-insensitive amplifier can be readily applied to other optomechanical systems which have been recently demonstrated, including optical~\cite{DongWang2012,HillPainter2012} and hybrid optical/microwave systems~\cite{AndrewsLehnert2014}, providing an essential link to create hybrid networks of otherwise incompatible quantum systems \cite{XiangNori2013,ClelandInterf,KurizkiSchmiedmayer2015}. Extending our scheme to multiple cavity modes creates a universal hub for \red{electro/optomechanical signals at the quantum level \cite{Zoller2012QIP,Marquardt2012QIP}, with high-gain, high-power amplification enabling interconnection of remote systems. At optical frequencies \cite{Painter2009Xtal}, the quantum limit of added noise should be accessible at room temperature. Given that in the present microwave experiment,} the added noise was limited by residual heating of the mechanical resonator, we expect the quantum limit can be reached by, first of all, improving the coupling efficiency. With a factor of two higher coupling, and an order of magnitude higher pump powers feasible in particular in 3D cavities \cite{Steele3D} with niobium technology, the device can operate at MHz-range bandwidth close to the standard quantum limit at 4 Kelvin temperatures, hence presenting an attractive alternative to HEMT amplifiers in narrow-band microwave measurements. 

\textbf{Acknowledgements} We thank Visa Vesterinen and Pasi L\"ahteenm\"aki for useful discussions. This work was supported by the Academy of Finland (contract 250280, CoE LTQ, 275245) and by the European Research Council (240387-NEMSQED, 240362-Heattronics, 615755-CAVITYQPD). The work benefited from the facilities at the Micronova Nanofabrication Center and at the Low Temperature Laboratory infrastructure.

\textbf{Author contributions} C.F.O.-K. carried out the practical work, analyzed data and wrote the paper. E.D. and J.-M.P. developed the device fabrication process. T.T.H. and F.M. developed the theory. M.A.S initiated and supervised the project.


\bibliographystyle{casnature}
\bibliography{AmplifierPaper}

\end{document}